 \documentstyle[twocolumn,aps,epsf]{revtex}
\begin{document}
\draft
\title{
Finite Size Scaling for Low Energy Excitations  in Integer 
Heisenberg Spin Chains
}
\author{
Shaojin Qin${}^1$, Yu-Liang Liu${}^1$,  Lu Yu${}^{1,2}$
}
\address{
${}^1$International Center for Theoretical Physics, P. O. Box 586, 
34100 Trieste, Italy
}
\address{
${}^2$Institute of Theoretical Physics, P. O. Box 2735, Beijing 
100080, P.R. China
}
\date{ \today }
\maketitle
\begin{abstract}
In this paper we study the finite size scaling for low energy 
excitations of $S=1$ and $S=2$  Heisenberg chains, using the 
density matrix renormalization group technique.  A crossover 
from $1/L$ behavior (with $L$ as the chain length) for medium 
chain length to $1/L^2$ scaling for long chain length is found 
for excitations in the continuum band as the length of the open 
chain increases.  Topological spin $S=1/2$ excitations are shown to 
give rise to the two lowest energy states for both open and 
periodic $S=1$ chains.  In periodic chains these two excitations 
are ``confined'' next to each other, while for open chains they 
are two free edge 1/2 spins.  The finite size scaling of the two 
lowest energy excitations of open $S=2$ chains is determined by 
coupling the two free edge $S=1$ spins.  The gap and correlation length 
for $S=2$ open Heisenberg chains are shown to be 0.082 (in units of 
the exchange $J$) and 47, respectively.
\end{abstract}


\narrowtext
%\tighten

Since Haldane pointed out the difference between integer and 
half integer Heisenberg spin chains\cite{hal}, the richness 
of the physics in the Heisenberg model has attracted intensive 
attention.  It was conjectured\cite{hal} that the ground state 
of integer spin chains has an exponentially decaying  correlation 
and a nonzero energy gap for excitations, in contrast to the 
ground state for half integer spin chains with a power law decay 
correlation and a zero energy gap. Like valence bond solid 
picture\cite{vbs}, free edge spins are expected to exist in both 
integer and half integer open Heisenberg spin chains with site 
spin $S>1/2$ \cite{tkedg}.  For $S=1$ spin chains, the field 
theory prediction\cite{field} for the continuum band and magnetic 
susceptibility of $S=1$ chains  agrees with experiment\cite{exprx}; 
the energy gap is found both numerically \cite{gol} and 
experimentally \cite{expgp}.  The numerical calculation on effects 
of free edge spins for $S>1/2$ chains\cite{edge1} also agrees 
with previous analytic studies on edge states\cite{vbs,tkedg}.  
For $S=1$ chains, the finite size 
scaling for energy is expected to be $1/L^2$ (L is the chain length) 
for the multi-magnon states with periodic boundary condition 
(PBC)\cite{field}, and the same $1/L^2$ scaling is also anticipated 
for open boundary condition (OBC) chains with one or several 
magnons\cite{aff}, while the first excitation energy scales 
exponentially for PBC chains\cite{gol}.  In contrast, different
scaling formula have been used in studying $S=2$ chains 
\cite{s2gp,jap,saclay,sung,cox}.  Therefore, a systematic study 
of finite size scaling for the energy spectrum of integer spin 
chains at various chain length is highly needed to obtain a 
precise gap value for $S=2$ spin chains.

In this paper, we study numerically the finite size scaling for 
low excitation energies in the continuum band for $S=1$ and $S=2$ 
Heisenberg open chains, which exhibits a  $\frac 1 L$ behavior 
for medium chain length and $\frac 1 {L^2}$ dependence for long 
chain length.  The crossover between medium length behavior and 
long length behavior is demonstrated by numerical results for 
$S=1$ open chains.  Such a crossover reconciles the numerical 
$\frac 1 L$ scaling found recently \cite{jap,saclay} for $S=2$ 
spin chains and the theoretical $\frac 1{L^2}$ scaling for chain 
length $L\to\infty$.  The finite size scaling of energies for 
the two lowest states of $S=1$ Heisenberg chains is given and 
topological spin 1/2 excitations are shown to be responsible 
for them. We show explicitly that the two spin 1/2 edge states 
are ``confined'' to each other in PBC chains.  The finite size 
scaling of the two lowest excitation energies in $S=2$ OBC chains 
is also determined by their free edge $S=1$ spins.  The gap and 
correlation length for $S=2$ Heisenberg chains are shown to be 
0.082(3) (in units of the exchange $J$) and 47(3), respectively.  
We also point out that a reasonable scaling for the first excited 
state (the bottom of the continuum band) of $S=2$ PBC chains is 
of exponential type in analogy with $S=1$ PBC chains.

We apply the density matrix renormalization group (DMRG) method
\cite{dmrg} to calculate the lowest energies for Heisenberg 
chains using the Hamiltonian:
\begin{equation}
H=\sum_{i=1}^{L-1} {\bf S}_i \cdot {\bf S}_{i+1} .
\end{equation}
(The above equation is for OBC. The Hamiltonian for PBC chains is 
obtained  by adding the term  ${\bf S}_L \cdot {\bf S}_1$ to Eq.(1).)  
We use the DMRG infinite chain algorithm to calculate the low 
energy spectrum for both OBC chains and PBC chains\cite{dmrg}.  
By keeping 300 optimized states, we calculate the energies 
$E_0^{P1}$ for the ground state and $E_1^{P1}$ for the first 
excited state in $S=1$ PBC chains up to length $L=36$.  The 
largest truncation errors in calculations are $9\times 10^{-8}$ 
for $E_0^{P1}$ and $5\times 10^{-7}$ for $E_1^{P1}$, respectively.  
By keeping 
200 optimized states, we calculate the energies $E_0^{O1}$ for the 
ground state and $E_1^{O1}$ for the first excited state of $S=1$ 
OBC chains up to length $L=42$.  The largest truncation errors in 
the calculations are $2\times 10^{-11}$ for $E_0^{O1}$ and 
$1\times 10^{-13}$ for $E_1^{O1}$.  By keeping 150 optimized 
states, we calculate also the energies $E_2^{O1}$, $E_3^{O1}$, 
and $E_4^{O1}$ for the low-lying states of total spin $S_z^{tot}=2$, 
3, and 4, respectively for $S=1$ OBC chains up to length $L=100$.  
$E_1^{O1}$ and $E_2^{O1}$ are then calculated for chain lengths up 
to $L=250$. The largest truncation errors occurred in these 
calculations are $1\times 10^{-12}$, $2\times 10^{-10}$, 
$1\times 10^{-9}$, and $6\times 10^{-9}$ for $E_1^{O1}$, 
$E_2^{O1}$, $E_3^{O1}$, and $E_4^{O1}$, respectively.  By keeping 
250 optimized states, we calculate the energies $E_0^{O2}$, 
$E_1^{O2}$, $E_2^{O2}$, and $E_3^{O2}$ for $S=2$ OBC chains up to 
length $L=150$, and they are corresponding to the lowest energy 
states with total spin $S_z^{tot}=0$, 1, 2, and 3, respectively.  
The largest truncation errors in the calculations are 
$1 \times 10^{-6}$, $2\times 10^{-7}$, $4\times 10^{-8}$, and 
$7\times 10^{-8}$ for $E_0^{O2}$, $E_1^{O2}$, $E_2^{O2}$, and 
$E_3^{O2}$, respectively.  The analysis of these numerical results 
is straightforward, and we will show them in detail in the following.

We use the following relations for the two lowest energies in $S=1$ 
Heisenberg chains (to be explained later):
\begin{equation}
\begin{array}{l}
E_0^{P1}=e_0 L + a_0^{P1} e^{-L/\xi_1}/\sqrt{L}, \\
E_1^{P1}=e_0 L+\Delta_1+a_1^{P1} e^{-L/\xi_1}/\sqrt{L},  \\
E_0^{O1}=e_0 (L-1)+\Delta_b+a_0^{O1} e^{-(L-1)/\xi_1}, \\
E_1^{O1}=e_0 (L-1)+\Delta_b
		-\frac{1}{3}a_0^{O1} e^{-(L-1)/\xi_1}.
\end{array}
\end{equation}
for PBC (the first two equations) and OBC (the last two relations),
respectively. It is more convenient to introduce 
\begin{equation}
\begin{array}{l}
f_1(L)=-(E_0^{P1} - e_0 L)\sqrt{L} 
	= -a_0^{P1} e^{-L/\xi_1}, \\
f_2(L)=(E_1^{P1} - e_0 L -\Delta_1)\sqrt{L} 
	= a_1^{P1} e^{-L/\xi_1},  \\
f_3(L \!\! - \!\! 1)= \!\! - 
	(E_0^{O1} \!\! - \! e_0 (L \!\! - \!\! 1)\!\!-\!\! \Delta_b)
	= -a_0^{O1} e^{-(L-1)/\xi_1},  \\
f_4(L \!\! - \!\! 1)= 
	E_0^{O1} \!\! - \! e_0 (L \!\! - \!\! 1)\!\!-\!\! \Delta_b
	= -\frac{1}{3}a_0^{O1} e^{-(L-1)/\xi_1}.
\end{array}
\end{equation}
%Fig.1
\begin{figure}[hbt]
\epsfxsize=\columnwidth\epsfbox{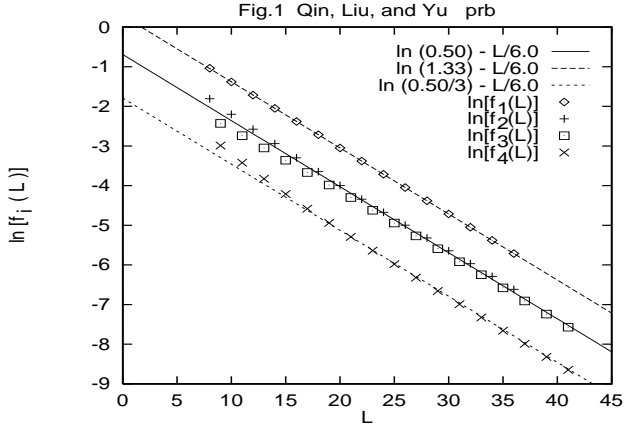}
\caption{
The logarithm of the energy expressions in Eq.(3) for the two lowest 
energy states of $S=1$ spin chains as functions of chain length L.
}
\end{figure}
Our numerical results are plotted in Fig.1 as 
$\ln f_i(L)$ v.s. $L$ in comparison  with Eq.(3).
From fitting we obtain the site energy $e_0=-1.401484(1)$, 
correlation length $\xi_1=6.00(4)$, energy gap 
$\Delta_1=0.4104892(2)$, and edge state energy for OBC chains
$\Delta_b=-0.1931661(1)$.  The constants are of similar precision 
for $\xi_1$, $a_0^{P1}=-1.33(1)$, $a_1^{P1}/a_0^{P1}=-0.33(2)$, 
and $a_0^{O1}=-0.50(2)$.

The interpretation of OBC results is obvious: $E_0^{O1}$ and 
$E_1^{O1}$ are the energies of the singlet and triplet combinations
of the two spin $1/2$ edge states, respectively.  Hence the
ratio of the constants is $-\frac 1 3$ and the coupling constant 
$J_{eff}\sim e^{-(L-1)/\xi}$ \cite{edge1,aff,white}. 
It is very interesting to note that $a_1^{P1}/a_0^{P1} \sim -1/3$ in 
Eq.(2),  which means that the singlet and triplet combinations of 
the two topological spin 1/2 excitations are also responsible for 
the two lowest states for PBC chains, with a coupling constant 
$J_{eff}\sim e^{-L/\xi}/\sqrt{L}$.  An important question is where 
are these topological excitations located.  For OBC chains they are 
exactly pinned at the edges. One might think that these two states 
are pushed away as far as possible from each other for PBC chains.  
However, this is not true: they are ``confined'', 
sitting next to each other\cite{afconf}.  
These two excitations will interact with each other in 
``both directions'', a direct coupling of the order of $J$ which 
is responsible for the Haldane gap, and an indirect coupling via 
virtual spin-wave excitations around the spin chain which is 
proportional to $J_{eff}\sim \exp(-L/\xi)/\sqrt{L}$\cite{ng1}. 

The exponential term $e^{-L/\xi}/\sqrt{L}$ in $E_0^{P1}$ and 
$E_1^{P1}$ is in agreement with previous exact diagonalization 
results\cite{gol}. By studying chain length longer than the one used 
in exact diagonalization, we have shown that $E_0^{P1}$ and $E_1^{P1}$ 
have the same scaling behavior as for the spin-spin correlation
\cite{gol,white}, instead of a net exponential $e^{-L/\xi}$.  We 
point out that $E_1^{P1}$ is the bottom of the one magnon band and has 
a zero kinetic energy since $w(k)\sim(k-\pi)^2$\cite{field,aff}, 
therefore its scaling is different from the $1/L^2$ behavior expected 
for other single or multi-magnon states in PBC chains as well as the 
magnon states in OBC chains.
%Fig.2
\begin{figure}[hbt]
\epsfxsize=\columnwidth\epsfbox{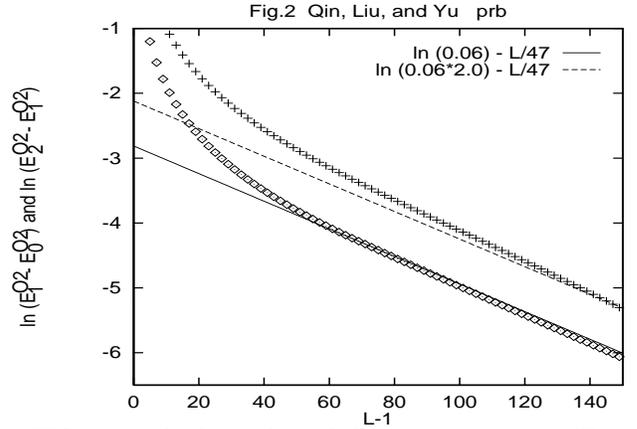}
\caption{
The logarithm of the energy spacings (Eq.(4)) for the three lowest 
energy states in $S=2$ spin chains as functions of chain length L.
Diamonds are for $\ln (E_1^{O2}-E_0^{O2})$, and crosses are for
$\ln (E_2^{O2}-E_1^{O2})$. 
}
\end{figure}
Now we turn to $S=2$ chains.  The asymptotic energy differences for 
the three lowest states in $S=2$ OBC chains are:
$$ E_1^{O2} - E_0^{O2}=a_0^{O2} e^{-(L-1)/\xi_2}, $$
\begin{equation}
E_2^{O2} - E_1^{O2}=2 a_0^{O2} e^{-(L-1)/\xi_2},
\end{equation}
where the correlation length $\xi_2=47(3)$ and constant 
$a_0^{O2}=0.060(5)$.  
The DMRG data are plotted in comparison with the above equation  in 
Fig.2 as $\ln(E_1^{O2}-E_0^{O2})$ and $\ln(E_2^{O2}-E_1^{O2})$ v.s. 
$L-1$.  These asymptotic expressions come from the coupling of two spin 
$S=1$ edge spins in $S=2$ OBC chains\cite{tkedg,edge1}.  From these 
energies we have calculated the site energy for $S=2$ Heisenberg chains
$\epsilon_0=-4.76125(5)$ by $E_0^{O2}(L+2)-E_0^{O2}(L)$ 
analysis\cite{white}.
%Fig.3
\begin{figure}[hbt]
\epsfxsize=\columnwidth\epsfbox{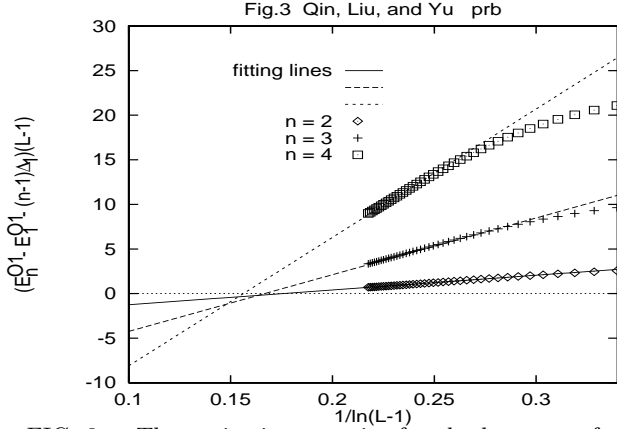}
\caption{
The excitation energies for the bottoms of one, two, and three magnon 
bands of medium length $S=1$ OBC chains (Eq.(5)).  We plot 
$[E_n^{O1}-E_1^{O1}-(n-1)\Delta_1](L-1)$ v.s. $1/\ln (L-1)$ for medium 
lengths $L=20$ to 100.
}
\end{figure}
%Fig.4
\begin{figure}[hbt]
\epsfxsize=\columnwidth\epsfbox{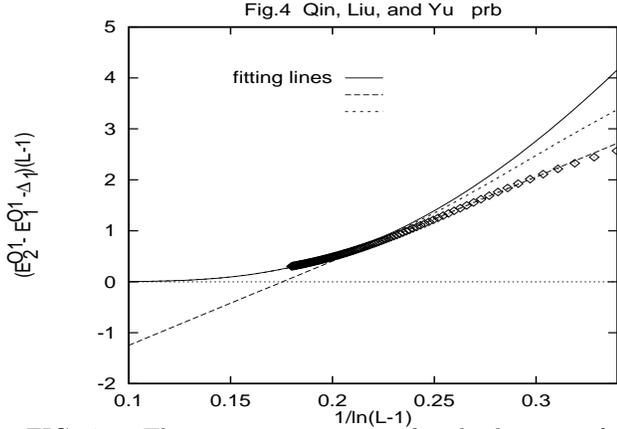}
\caption{
The excitation energies for the bottom of one magnon band of 
medium length and long length $S=1$ OBC chains.  We plot 
$(E_2^{O1}-E_1^{O1}-\Delta_1)(L-1)$ v.s. $1/\ln (L-1)$ for medium 
lengths $L=20$ to 100, and for long lengths $L =100$ to 250.  The 
crossover from $(E_2^{O1}-E_1^{O1}-\Delta_1)(L-1)$ $\approx$ 
$a_2^{O1}(1+b_2^{O1}/\ln(L-1))$ for medium lengths (dashed line) to
$(E_2^{O1}-E_1^{O1}-\Delta_1)(L-1)$ $=$ $74.7/ (L-1)^2$ for long 
lengths (full line) {\protect \cite{aff}} is evident.  The bottom 
of the magnon dispersion $E_2^{O1}-E_1^{O1}$ $=$ 
$\protect\sqrt{\Delta^2+(2.49\pi /L)^2}$ {\protect \cite{aff}} is 
also plotted (dotted line).
}
\end{figure}

As for excitations in the continuum, we first describe the medium 
length scaling behavior for integer spin OBC chains. For $S=1$ 
chains,  we use the following energy expression for medium length:
\begin{equation}
E_n^{O1}-E_1^{O1}=(n-1)\Delta_1+ 
	{{a^{O1}_n} \over {L-1}}(1+{{b^{O1}_n}\over \ln (L-1)}),
\end{equation}
where $n=2, 3, 4$ correspond to one, two, and three magnon band 
bottoms, respectively.  We plot in Fig.3 
$(L-1)(E_2^{O1}-E_1^{O1} -\Delta_1)$, 
$(L-1)(E_3^{O1}-E_1^{O1} -2\Delta_1)$, and 
$(L-1)(E_4^{O1}-E_1^{O1} -3\Delta_1)$ v.s. $1/\ln (L-1)$ for 
$S=1$ OBC chains with medium chain length, $L=20$ to $100$, which 
is more than ten times the correlation length $\xi_1$.  The 
ratios of the constants are $a^{O1}_3/a^{O1}_2 \approx 4$,
$a^{O1}_4/a^{O1}_2 \approx 9$, and 
$b^{O1}_4 \approx b^{O1}_3 \approx b^{O1}_2$.  For long chains, 
$L>100$, the expected scaling term $1/L^2$ \cite{field,aff} for 
$E_n^{O1}-E_1^{O1}$ shows up.  In Fig.4, using again the plot 
$(L-1)(E_2^{O1}-E_1^{O1} -\Delta_1)$ v.s. $1/\ln (L-1)$ for chain 
length $L=20$ to $250$, we show the crossover from the behavior 
given by Eq.(5) for medium chain length to 
$E_2^{O1}-E_1^{O1}=\Delta_1 +\frac {74.7}{(L-1)^2}$ \cite{aff} for 
long chain length.  We point out that the crossover occurs within
a rather narrow range in $1/\ln L$ plot in Fig.4.
For $S=2$ OBC chains, we use medium chain length expressions 
similar to the ones for $S=1$ chains,
\begin{equation}
E_3^{O2}-E_2^{O2}=\Delta_2+ 
	{{a^{O2}_3} \over {L-1}}(1+{{b^{O2}_3}\over \ln (L-1)}).
\end{equation}
where the gap turns out to be $\Delta_2=0.082(3)$ and constants 
$a_3^{O2}=2.8(2)$ and $b_3^{O2}=5.9(4)$.  We plot in Fig.5 
$(L-1)(E_3^{O2}-E_2^{O2} -\Delta_2)$ v.s. $1/\ln(L-1)$ for chain 
length up to $L=150$.  As we show in Fig.5, the medium chain 
length behavior extends to chain length $L=150$ and bigger.  For 
the chain lengths we have studied, which is more than three times 
the correlation length, the leading term $1/L$ in Eq.(6) is 
still dominant.  In analogy to $S=1$ chains\cite{field,aff}, the 
long length asymptotic expression for $S=2$ OBC chains is expected 
to be $\frac 1 {L^2}$ as well\cite{saclay}, and the crossover 
between the medium length and long length behavior should occur 
for longer chain lengths.

One way to explain the $1/L$ scaling in the medium range is to consider 
the energy dispersion\cite{spc1} 
$\Delta(k)=\sqrt{\Delta^2+(vk)^2}$ for a magnon with wave 
vector $k$.  This expression has also been suggested for such 
gapped systems using nonlinear sigma models \cite{spc2}.  
For OBC chains with big length $L$, one can expand
the $(vk)^2$ term and get the $1/L^2$ scaling.  So it is in 
agreement with our calculation and previous studies\cite{aff}.  
The leading finite size scaling term would be $1/L$ only when 
$\Delta^2  \ll (2\pi v/L)^2$.  The shape of the curve 
$\sqrt{\Delta^2+(2\pi v/L)^2}$ v.s. $L$ 
is consistent with $E_2^{O1}-E_1^{O1}$ (see Fig.4), 
but it can not fit $E_2^{O1}-E_1^{O1}$ for the medium 
size well, for instance, $L\sim 20$ to $90$ for $S=1$ OBC chains.  
We suggest that further studies on the medium size behavior of 
$S=1$ and $S=2$ chains are needed, especially because it is relevant to 
experiment on chains with typical doping (of the order of one percent).
%Fig.5
\begin{figure}[hbt]
\epsfxsize=\columnwidth\epsfbox{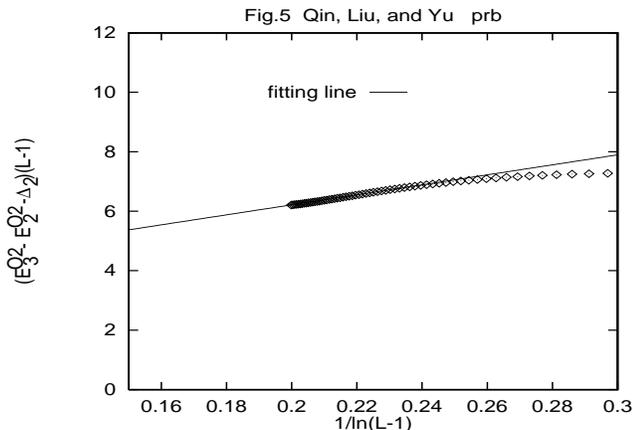}
\caption{
The excitation energies for the bottom of one magnon band of
medium length $S=2$ OBC chains (Eq.(6)).  We plot 
$(E_3^{O2}-E_2^{O2}-\Delta_2)(L-1)$ v.s.  $1/\ln (L-1)$ for lengths
$L =50$ to 150. 
}
\end{figure}

The leading order scaling $1/L$ for medium chain length has been found 
numerically in recent studies\cite{jap,saclay} for $S=2$ Heisenberg 
chains under various boundary conditions.  Given the decreasing of the 
gap equal to, or more rapid than, the inverse of chain length $1/L$, 
our calculation gives a lower bound $0.082(3)$ for the gap of $S=2$ 
chain.  In a previous work\cite{saclay}, a similar lower bound 
$0.085(5)$ was also given for the gap using a different boundary 
condition\cite{s1bd,lowb}.  Since the lower bound of the gap is  
finite, the existence of a nonzero gap in $S=2$ Heisenberg chain 
has been proved by numerical calculations, and therefore, Haldane's 
conjecture is verified also for $S=2$ chain.  Moreover, 
the calculated gap is also in agreement  with 
an earlier calculation in Ref.\cite{cox} and a recent experiment 
\cite{flor}.  We emphasize that the gap calculation in the present 
paper disagrees with some other studies\cite{s2gp,jap}.  This 
discrepancy may arise from the different scalings used, such as 
$1/L^2$ scaling \cite{sung} for the gap of short $S=2$ PBC chain, 
where an exponential type scaling is more appropriate.

To conclude we have performed a systematic study of finite size
scaling behavior for the energy spectrum of integer spin chains
($S=1,2$), using the DMRG technique.  This analysis shows that
the lowest excitations are due to the ``edge'' states for both OBC
and PBC chains.  As for excitations in the continuum a crossover 
from $1/L$ behavior for medium chain length to $1/L^2$ behavior for 
long chain length has been demonstrated.  Using this analysis, a 
reliable estimate for the gap and correlation length in $S=2$ chains
has been obtained.

%\acknowledgements
We would like to thank Profs. I. Affleck, T.K. Ng, and Z.B. Su 
for very helpful discussions.
The calculation was completed on IBM RISC-6000 at 
International Center for Theoretical Physics, Trieste.


\begin{references}
\bibitem{lowb} In \cite{saclay}, a rather peculiar boundary condition
		has been used and the authors have argued that their 
		boundary condition is appropriate.  From another point 
		of view, the boundary acts as an impurity for the chain
		and thus the lowest excitation energy should be less 
		than or equal to the Haldane gap of the chain in 
		certain situations\cite{white,s1bd}. So the lowest 
		excitation energy is the lower bound of the Haldane 
		gap in such systems.
\end{references}
\end{document}